\documentclass{aa}
\usepackage{amsmath}
\usepackage{graphicx}
\usepackage{txfonts}

\usepackage{natbib}
\bibpunct{(}{)}{;}{a}{}{,}

\newcommand{\unit}[1]{\ensuremath{\:\mathrm{#1}}}

\begin{document}

\title{Gravity modes in rapidly rotating stars}
\subtitle{Limits of perturbative methods}

\author{J. Ballot\inst{1}
\and  F. Ligni\`eres\inst{1}
\and  D. R. Reese\inst{2}
\and  M. Rieutord\inst{1}}

\institute{
Laboratoire d'Astrophysique de Toulouse-Tarbes, Universit\'e de Toulouse, 
CNRS, 14 avenue E. Belin, 31400 Toulouse, France\\
\email{jballot@ast.obs-mip.fr}
\and
LESIA, UMR8109, Universit\'e Pierre et Marie Curie, Universit\'e Denis Diderot, Observatoire de Paris, 92195 Meudon, France}

\date{Received 13 March 2010 / Accepted 3 May 2010}

\abstract%
{CoRoT and Kepler missions are now providing high-quality asteroseismic data for a large number of stars. Among intermediate-mass and massive stars, fast rotators are common objects. Taking the rotation effects into account is needed to correctly understand, identify, and interpret the observed oscillation frequencies of these stars. A classical approach is to consider the rotation as a perturbation.}%
{In this paper, we focus on gravity modes, such as those occurring in $\gamma$~Doradus, slowly pulsating B (SPB), or Be stars. We aim to define the suitability of perturbative methods.}%
{With the two-dimensional oscillation program (TOP), we performed complete computations of gravity modes -- including the Coriolis force, the centrifugal distortion, and compressible effects -- in 2-D distorted polytropic models of stars. We started with the modes $\ell=1$, $n=1$--14, and $\ell=2$--3, $n=1$--5, 16--20 of a nonrotating star, and followed these modes by increasing the rotation rate up to 70\% of the break-up rotation rate. We then derived perturbative coefficients and determined the domains of validity of the perturbative methods.}%
{Second-order perturbative methods are suited to computing low-order, low-degree mode frequencies up to rotation speeds $\sim$100\unit{km\,s^{-1}} for typical $\gamma$~Dor stars or  $\sim$150\unit{km\,s^{-1}} for B stars. The domains of validity can be extended by a few tens of \unit{km\,s^{-1}} thanks to the third-order terms. For higher order modes, the domains of validity are noticeably reduced. Moreover, perturbative methods are inefficient for modes with frequencies lower than the Coriolis frequency $2\Omega$. We interpret this failure as a consequence of  a modification in the shape of the resonant cavity that is not taken into account in the perturbative approach.}%
{}

\maketitle

\section{Introduction}\label{Sec:Intro}

CoRoT \citep[Convection, Rotation and planetary Transits,][]{BaglinM06_COROT} and Kepler \citep{BoruckiK07_KEPLER} are space missions providing uninterrupted high-quality photometry time series over several months or years ideally suited for asteroseismic study. Asteroseismology provides very accurate determinations of the stellar parameters (mass, radius, age, etc.) and probes stellar structure to constrain physical processes occurring in stars.
The first step towards this goal requires correctly understanding the structure of the observed oscillation spectra, and especially correctly identifying the observed modes. In the case of main-sequence \citep[e.g.][]{Michel08,Benomar09,Chaplin10} and giant  \citep[e.g.][]{Miglio09,Hekker09,Bedding10} FGK stars with solar-like oscillations, the spectrum structure is well understood, which eases interpretation.

The spectra of classical pulsators is often noticeably more complex.
For instance, the high-quality observations of $\delta$ Scuti \citep[e.g.][]{GarciaHernandez09,Poretti09} and $\gamma$~Doradus \citep[e.g.][]{Mathias09} stars have exhibited very rich and complex spectra of acoustic (p) and gravity (g) modes, respectively, containing several hundred --or more-- modes.
Interpreting their spectra is very challenging today. Indeed, these stars generally spin rapidly, so the effects of rotation on the mode frequencies must be considered.

Here, we are concerned with gravity modes, i.e., low-frequency modes driven by the buoyancy force. 
They are excited and observed in a broad panel of stars, for instance, in $\gamma$~Dor, SPB and some Be stars.
The $\gamma$~Dor stars form a class of main-sequence stars with type around F0V that can sometimes rotate rapidly \citep[e.g.][and reference therein]{DeCat06}, while the rotation rate of Be stars is extreme, usually very close to their break-up limit $\Omega\approx\Omega_K\equiv\sqrt{GM/R^3}$ \citep[e.g.][]{Fremat06}.

The effects of rotation on the oscillation modes can be treated as a perturbation where the rotation rate is 
the small parameter. A 1st-order correction has been proposed by \citet{Ledoux51}, 2nd-order by \citet{Saio81}, \citet{DziembowskiGoode92}, or \citet{Suarez06},  and 3rd-order terms have been developed by \citet{Soufi98}.
While perturbative methods are expected to be accurate enough for slowly rotating stars, 
their true domain of validity cannot be determined in the absence of exact calculations to compare them with.

In the past few years, calculations of p~modes with both the centrifugal distortion and the Coriolis force have been performed in polytropic models of stars \citep{Lignieres06,Reese06} and realistic 2-D stellar structures \citep{Lovekin08, Reese09}. 
\citet{Lignieres06} and \citet{Reese06} have shown that,
above $\Omega \sim 0.15\Omega_K$,
perturbation methods fail to reproduce low-degree and low-order p-mode frequencies ($\ell \le 3$ and $n \le 10$) with
the accuracy of CoRoT long runs. 
The structure of the modes is also drastically modified,
and this leads to deep changes in the structure of the p-mode spectrum \citep{Reese08,Lignieres08,Lignieres09}.

 We used an oscillation code based on \citet{Reese06} to perform
g-mode calculations with a complete description of the rotational effects on the modes.
In this paper, we focus on the limits of validity for perturbative methods.
The models and the method are described in Sect.~\ref{Sec:Methods}.
We then derive the perturbative coefficients from 
the complete computations (Sect.~\ref{Sec:Perturbative}), and compare the results obtained with
both methods to determine and discuss the domains of validity for perturbative methods (Sect.~\ref{Sec:Validity}) 
before concluding in Sect.~\ref{Sec:Concl}.

\section{Models and methods}\label{Sec:Methods}
We consider fully radiative stars for this work. Since the gravity modes are driven by the buoyancy force, they cannot exist in convective regions. SPB and $\gamma$~Dor stars have large radiative zones with a convective core, and even a thin convective envelope for the latter. The effects of convective cores are not considered here, since we are mainly interested in the general behavior of g modes under rotation effects.

\subsection{2-D stellar models}\label{SSec:Model}
As in \citet{Lignieres06} and \citet{Reese06}, we approximate the equilibrium structure of rotating stars with self-gravitating uniformly-rotating polytropes. They are described in the co-rotating frame by the three following equations:
\begin{eqnarray}
p_o&=&K\rho_o^{1+1/\mu} \\
\vec\nabla p_o &=& \rho_o \vec{g}_o \\
\Delta \psi_o &=& 4\pi G \rho_o
\end{eqnarray}
where $p_o$ is the pressure, $\rho_o$ the density, $\psi_o$ the gravitational potential, $K$ the polytropic constant, $\mu$ the polytropic index, $G$ the gravitational constant, and $\vec g_o$ the effective gravity defined as
\begin{equation}
\vec{g}_o = - \vec{\nabla} ( \psi_o - \Omega^2 s^2 /2)
\end{equation} 
with $s$ the distance to the rotation axis.
Due to the centrifugal distortion, the star is not spherical and a suited surface-fitting spheroidal system of coordinates $(\zeta,\theta,\phi)$ based on \citet{Bonazzola98} has been used. Hereafter, we also classically denote $r$ the distance to the center, and $z$ the coordinate along the rotation axis. This equation system is numerically solved with the ESTER (Evolution STEllaire en Rotation) code as described in \citet{Rieutord05}. This is a spectral code using Chebychev polynomials in the $\zeta$-direction, and spherical harmonics $Y_{\ell}^{m}$ with even $\ell$ and $m=0$ in the horizontal one.
We computed models decomposed on spherical harmonics up to degree $L_{\mathrm{model}}=32$. This resolution is high enough to accurately model the centrifugal effects for the maximal value of $\Omega$ that we have considered. In the pseudo-radial direction, the resolution is the same as the one we use for the frequency computation (see Sect.~\ref{ssec:resol}, $n_r=96$ generally).

To approximate a fully radiative star, we chose the polytropic index $\mu=3$. 
We considered models spinning with rotation frequency $\Omega$ between $0$ and $0.7\Omega_K$, where 
$\Omega_K=\sqrt{GM/R_{eq}}$ is the Keplerian break-up rotation rate for a star of mass $M$ and equatorial radius $R_{eq}$.

\subsection{Linearized equations for the oscillations}
In the co-rotating frame, the equations governing the temporal evolution of small adiabatic inviscid perturbations of the equilibrium structure read, in the co-rotating frame,
\begin{eqnarray}
\partial_t \rho &=& -\vec\nabla\cdot (\rho_o \vec v) \label{eq:pert1}\\
\rho_o \partial_t \vec v &=&  -\vec\nabla p + \rho \vec g_o -\rho_o \vec\nabla \psi - 2\rho_o \vec\Omega \times \vec v\\
\partial_t p - c_o^2 \partial_t \rho &=& \frac{\rho_oN_o^2c_o^2}{||\vec{g}_o||^2} \vec v\cdot \vec g_o \label{eq:pertenerg}\\
\Delta \psi &=& 4\pi G \rho \label{eq:pert4}
\end{eqnarray}
where $\rho$, $p$, $\vec{v}$, and $\psi$ are the Eulerian perturbations of density, pressure, velocity, and gravitational potential
$c_o^2=\Gamma_1 p_o /\rho_o $ the adiabatic sound speed and $N_o$ the Brunt-V\"ais\"al\"a frequency, defined as
\begin{equation}
  \label{eq:BV}
  N_o^2=\vec g_o \cdot \left( \frac{\vec\nabla \rho_o}{\rho_o} -\frac{1}{\Gamma_1}\frac{\vec\nabla p_o}{p_o}\right).
\end{equation}
$\Gamma_1=(\partial\ln p /\partial\ln\rho)_{\mathrm{ad}}$ denotes the first adiabatic exponent.
In Eq.~(\ref{eq:pertenerg}) we have used the structure barotropicity, ensured by the uniform rotation.

\begin{figure}[!tp]
  \begin{center}
    \hspace{.1\linewidth}\includegraphics[width=.9\linewidth]{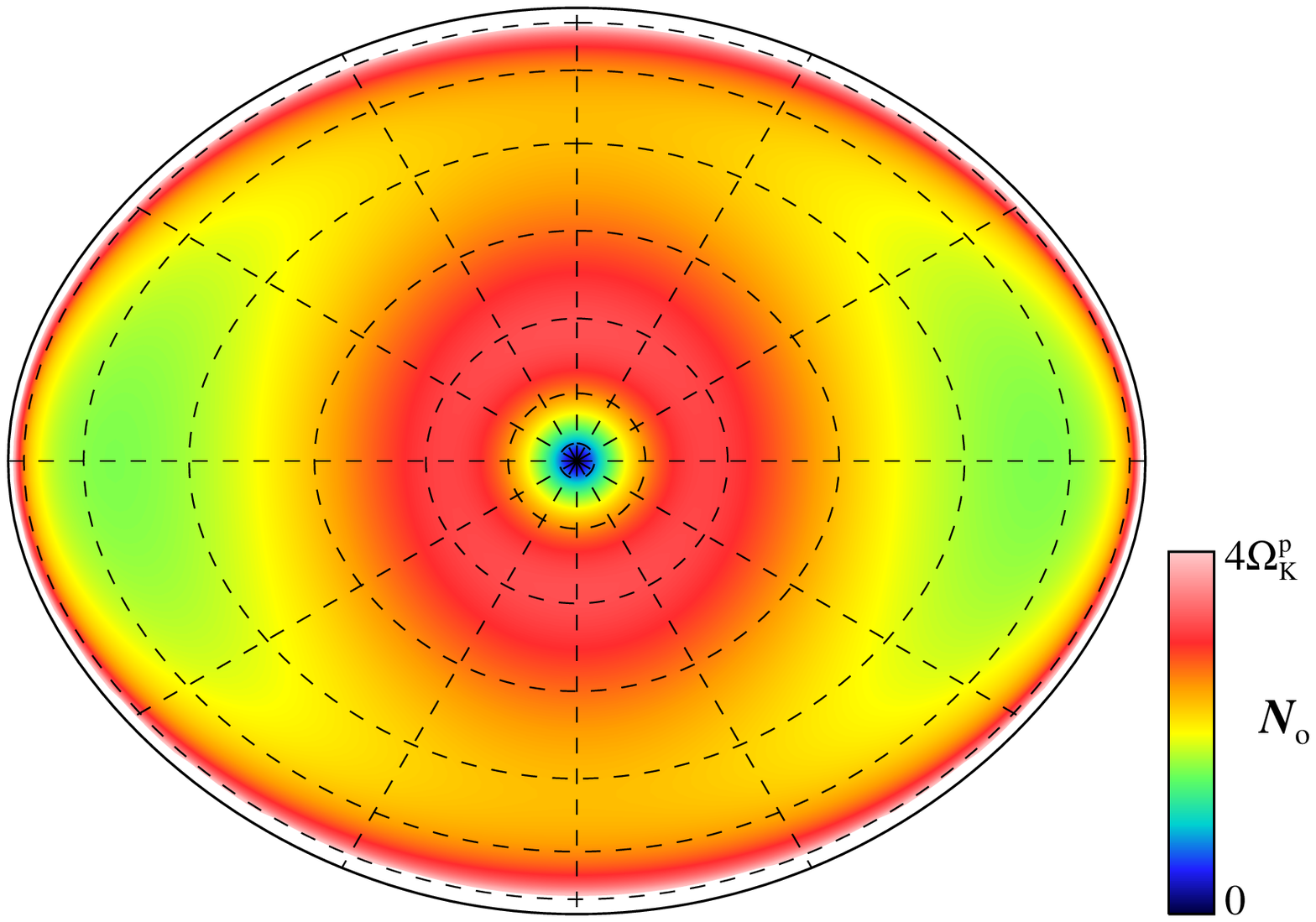}\\
    \includegraphics[width=\linewidth]{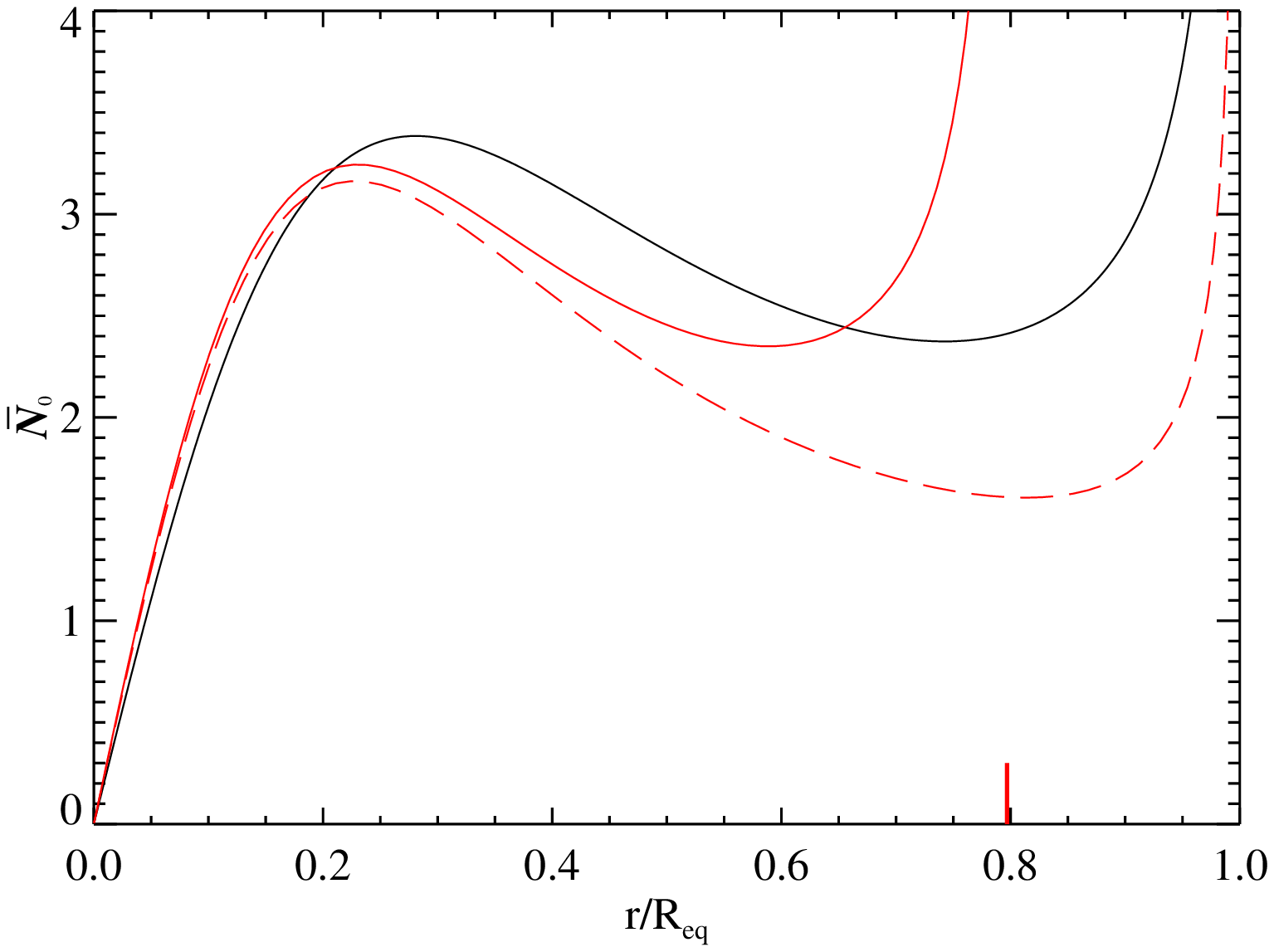}
  \end{center}
  \caption{\textit{(top)} Map in the meridional plane of the Brunt-V\"ais\"al\"a frequency $N_o$ for the model with $\Omega=0.7\Omega_K$. $N_o$ has been normalized by $\Omega_k^p=\sqrt{GM/R_{p}}$ where $R_{p}$ is the polar radius. Dashed lines show the shape of the spheroidal grid we used.
\textit{(bottom)} Solid black lines show the profile of $N_o$ normalized by $\Omega_K^p$ for the nonrotating star. Red lines show $N_o$ for the model $\Omega=0.7\Omega_K$ along the polar (solid line) and equatorial radius (dashes). A thick red tick on the x-axis indicates the polar radius $R_p$. }
     \label{fig:BV}
   \end{figure}

The 2-D distribution of the Brunt-V\"ais\"al\"a frequency is shown in Fig.~\ref{fig:BV} for the most rapidly rotating model we have considered, together
with the profiles of $N_o$ along the polar and equatorial radii, which are compared to the $N_o$ profile of the nonrotating star.
Within about the inner half of the star, the deviations from sphericity induced by the centrifugal 
force remain limited.
We also see that $N_o$ diverges at the surface of the polytrope because $\rho_o$ and $p_o$ vanish there. 

Looking for time-harmonic solutions $\propto \exp (i\omega t)$ of the system (\ref{eq:pert1})--(\ref{eq:pert4}), we obtain an eigenvalue problem, which we then solve using the two-dimensional oscillation program (TOP).
The details of this oscillation code closely follow \citet{Reese06}.
The equations are projected on the spherical harmonic basis $Y_{\ell}^m$. Due to the axisymmetry of the system, the projected equations are decoupled relatively to the azimuthal order $m$, but in contrast to the spherical non-rotating case, they are coupled for all degrees $\ell$ of the same parity.

\subsection{Resolutions and method accuracy}\label{ssec:resol}

The different sources of error of our numerical method have been discussed in \citet{Valdettaro07} and tested in a context similar to the present one in \citet{Lignieres06} and \citet{Reese06}. 
The numerical resolution has been chosen to ensure a sufficient accuracy for the computed frequencies.
In the horizontal direction, the resolution is given by the truncation of the spherical harmonics expansion.
The highest degree of the expansion is $L=2n_{\theta}+|m|$ and,  for most of the calculations presented here, we used $L=40+|m|$, i.e. $n_\theta=20$ coupled spherical harmonics.
In the pseudo-radial direction $\zeta$, the solution has been expanded over the set of Chebychev polynomials up to $n_r=96$. 

\begin{figure}[!htp]
  \centering
  \includegraphics[width=\linewidth]{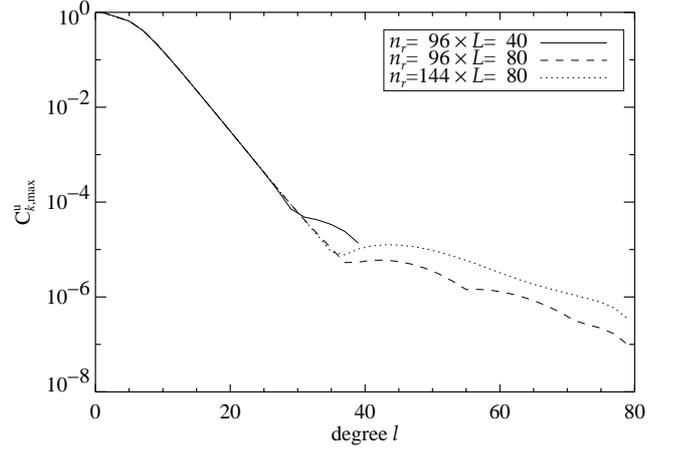}
  \caption{Spectrum $C^u_{k,max}$ as a function of the degree $\ell$ of the radial velocity of the mode $(\ell,m,n)=(1,0,14)$ at $\Omega=0.7\Omega_K$. $C^u_{k,max}(\ell)$ is the absolute value of the highest Chebychev coefficient in the decomposition on the spherical harmonics of degree $\ell$ of the radial component of $\vec v$. The spectrum is normalized to its maximum value. The different lines correspond to different spatial resolutions.}
  \label{fig:spectrum}
\end{figure}

Using higher resolutions ($n_r=96$, $L=80$ and $n_r=144$, $L=80$), we find that the relative agreement of the frequencies always remains better than $5\times 10^{-8}$. 
It also does not affect the mode significantly as illustrated in Fig.~\ref{fig:spectrum} where the spectral expansion of the radial velocity component of the $(\ell,m,n)=(1,0,14)$ mode at $\Omega=0.7\Omega_K$ is displayed for the three different resolutions. Figure~\ref{fig:spectrum} also shows that a unique --or even a few-- spherical harmonics would not properly describe such an eigenmode.

\subsection{Following modes with rotation}
We computed the frequencies of $\ell=1$ to 3 modes in a nonrotating polytrope. We recall that without rotation the system to solve becomes decoupled with respect to $\ell$, hence the modes are represented with only one spherical harmonic. A reference frequency set, $\omega_{\ell,n}^{(0)}$, was computed from a 1-D polytropic model with a radial resolution $n_r=512$.

We then followed the variation in frequency of each mode of degree $\ell_0$, azimuthal order $m_0$, and radial order $n_0$ by slowly increasing the rotation rate, step by step.
The Arnoldi-Chebychev method requires an initial guess for the frequency, and returns the solutions that are the closest to this guess.
The guess we provide is extrapolated from the results at lower rotation rates: we compute from the three last computed points a quadratic extrapolation at the desired rotation rate. For the first point ($\Omega=0$), we use the frequency obtained in 1-D as a guess. For the second point, we extrapolate a guess with the asymptotic relation $\omega_{\ell_0,m_0,n_0}\approx\omega_{\ell_0,m_0}^{(0)}+m_0\Omega/[\ell_0(\ell_0+1)]$ \citep{Ledoux51}.

Among the solutions found around the initial guess, we select the correct one by following this strategy:
\begin{enumerate}
\item For each calculated mode, we determine, from its spatial spectrum (like the ones shown Fig.~\ref{fig:spectrum}), the two dominant degrees, $\ell_1$ and $\ell_2$.
\item We compare $\ell_1$ and $\ell_2$ with the degree $\ell_0$ of the mode we are following.
\item We select the solutions such that $\ell_1=\ell_0$; if none of the solutions verifies this criterion, we select the solutions such that $\ell_2=\ell_0$.
\item If more than one solution has been selected at this point, we consider the projection of the modes on the spherical harmonic $Y_{\ell_0}^{m_0}$ and compare it to the projection of the mode at a lower rotation rate. The solution that gives the highest correlation is finally selected.
\end{enumerate}

This method allows us to have a semi-automatic procedure that limits the need to manually tag the modes. Nevertheless, there are two limitations. The first one is inherent to the density of the g-mode spectrum. Indeed, according to the asymptotic relation from \citet{Tassoul80}, on a given frequency interval, the number of modes of degree $\ell$ scales roughly as $\sqrt{\ell(\ell+1)}$. If our resolutions, both in latitude and radius, were infinite, it would almost be impossible to find the desired solution, beacause there would always be an infinite number of other solutions closer to the initial guess than the mode being searched. In practice, the spatial resolution is finite, and we noticed that, in most cases, the number of solutions in a small interval is limited enough to allow us to find the desired solution. Another way of avoiding this difficulty is to add thermal dissipation that disperses the different solutions in the complex plane. This also proved successful in finding a specific solution.

\begin{figure}[!htp]
  \centering
  \includegraphics[width=\linewidth]{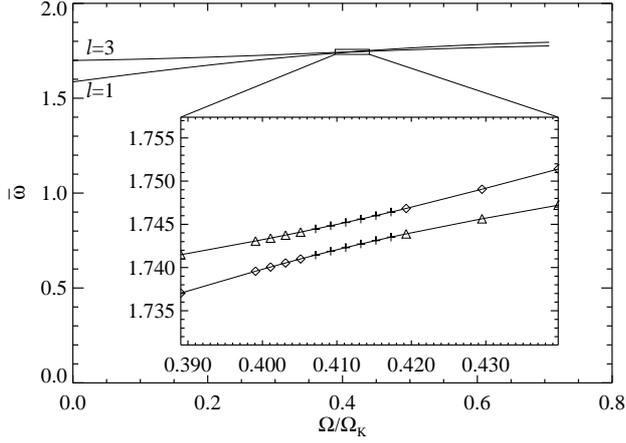}
  \caption{Evolution with the rotation rate of the frequency of the modes $(\ell=1,m=1, n=1)$ and $(\ell=3, m=1, n=3)$, with a zoom on a region where an avoided crossing between these two modes occurs. At a given rotation rate, a diamond (respectively, a triangle) indicates the mode is dominated by the component $\ell=1$ ($\ell=3$). Crosses indicate that the modes are hardly discernible: they are both dominated by $\ell=1$ and their structures are very similar.}
  \label{fig:cross}
\end{figure}

The second difficulty comes from the so-called avoided crossings. Two modes with the same $m$ and the same parity cannot have the same frequency. This implies that the two curves associated to their evolution with $\Omega$ cannot cross each other. Figure~\ref{fig:cross} illustrates this phenomenon with modes $\ell=1$ and $\ell=3$: the frequencies get closer and closer, but since the curves cannot cross, the modes exchange their properties. During an avoided crossing, the two modes have the mixed properties of the two initial modes.
With our mode-following method, when the coupling is strong and the avoided crossing takes long, the method can follow the wrong branch. For instance, in the case illustrated in Fig.~\ref{fig:cross}, if the program follows the $\ell=1$ mode, it continues sometimes on the $\ell=3$ branch instead of jumping to the other branch.

\section{Perturbative coefficients}\label{Sec:Perturbative}

The approach used to determine the perturbative coefficients in this paper is very close to the one of \citet{Reese06}.
\subsection{Determining perturbative coefficients}
In the perturbative approach, frequencies are developed as a function of the rotation rate, $\Omega$. For instance, to the 3rd order, it reads
\begin{equation}
\bar\omega_{\ell,m,n}^{pert}=\bar\omega_{\ell,n}^{(0)}+ C_{\ell,m,n}^1 \bar\Omega + C_{\ell,m,n}^2 \bar\Omega^2 + C_{\ell,m,n}^3 \bar\Omega^3 + {\cal O}(\bar\Omega^4),\label{eq:pertraw}
\end{equation}
where $\omega_{\ell,n}^{(0)}$ is the frequency for the non-rotating case, and $C_{\ell,m,n}^j$ the perturbative coefficients.
The bar denotes the normalization $\bar\omega=\omega/\Omega_K^{p}$ and $\bar\Omega=\Omega/\Omega_K^{p}$. We normalize the frequencies by $\Omega_K^{p}$ since the polar radius is expected to be a slowly varying function of $\Omega$ in real stars, as opposed to $R_{eq}$.

The coefficients $C_{\ell,m}^j$ can be numerically calculated from the complete computations since they are directly linked to the $j$-th derivative of the function $\bar\omega_{\ell,m,n}(\bar\Omega)$ at $\Omega=0$. However, to improve the accuracy, we use symmetry properties of the problem: changing $\Omega$ in $-\Omega$, one easily shows that
\begin{equation}
C_{\ell,-m,n}^j=(-1)^jC_{\ell,m,n}^j\qquad \forall m.
\end{equation}
We define
\begin{eqnarray}
x&=&\bar\Omega^2,\\
y^D_{\ell,m,n} &=& \frac{\bar\omega_{\ell,m,n}-\bar\omega_{\ell,-m,n}}{2\Omega}\qquad m>0,\\
y^S_{\ell,m,n} &=&\frac{1}{x}\left[\frac{\bar\omega_{\ell,m,n}+\bar\omega_{\ell,-m,n}}{2}-\bar\omega_{\ell,n}^{(0)}\right]\qquad m\geq 0,\label{eq:ySdef}
\end{eqnarray}
and get
\begin{eqnarray}
y^D_{\ell,m,n} &=& C_{\ell,m,n}^1 + C_{\ell,m,n}^3 x +  
 \cdots + C_{\ell,m,n}^{2k+1} x^k +{\cal O}(x^{k+1})\label{eq:yD}\\
y^S_{\ell,m,n} &=& C_{\ell,m,n}^2 + C_{\ell,m,n}^4 x +  
\cdots + C_{\ell,m,n}^{2k+2} x^k + {\cal O}(x^{k+1}).\label{eq:yS}
\end{eqnarray}
We note that $C_{\ell,m=0,n}^j$ vanish for odd $j$.

We compute $y^D_{\ell,m,n}$ and $y^S_{\ell,m,n}$ on a grid of $k$ points from $\Omega=\delta\bar\Omega$ to $k\delta\bar\Omega$ and use the Eqs.~(\ref{eq:yD}) and (\ref{eq:yS}) to calculate the terms $C_{\ell,m,n}^j$ with the $(k-1)$-th-order interpolating polynomials. The determination of the coefficients $C_{\ell,m,n}^j$ is then accurate to the $(2k-1)$-th-order in $\bar\Omega$. In practice we use a typical step $\delta\bar\Omega=2\times 10^{-3}$ and $k=4$.
In Eq.~(\ref{eq:ySdef}), we use $\bar\omega_{\ell,n}^{(0)}=[\bar\omega_{\ell,m,n}(\Omega=0)+\bar\omega_{\ell,-m,n}(\Omega=0)]/2$, making it totally independent of the 1-D solutions.

By explicitly expressing the dependence on $m$ of the perturbative coefficients, Eq.~(\ref{eq:pertraw}) becomes
\begin{multline}
\bar\omega_{\ell,m,n}^{pert}=\bar\omega_{\ell,n}^{(0)}+ mC_{\ell,n} \bar\Omega + 
(S^1_{\ell,n}+m^2S^2_{\ell,n}) \bar\Omega^2 + \\
m(T^1_{\ell,n}+m^2T^2_{\ell,n}) \bar\Omega^3 + {\cal O}(\bar\Omega^4) \label{eq:pertfin}
\end{multline}
The form of the 1st order comes from \citet{Ledoux51}, the 2nd order from \citet{Saio81}, and the 3rd is derived from \citet{Soufi98}. We have verified that the derived coefficients fit these relations with a very good accuracy (see below) and list them in Table~\ref{tab:coef}.

\begin{table*}[htp]
  \centering  
  \caption{Perturbative coefficients (see development Eq.~\ref{eq:pertfin}) for g modes with frequency $\bar\omega>0.255$, radial order $n\leq 25$, and $\ell \leq 3$ in a polytropic stellar model with an index $\mu=3$.}
  \label{tab:coef}
  \begin{tabular}{rrrrrrr}
\hline
\hline
\multicolumn{1}{c}{$n$}&
\multicolumn{1}{c}{$\bar\omega^{(0)}_{\ell,n}$}&
\multicolumn{1}{c}{$C_{\ell,n}$}&
\multicolumn{1}{c}{$S^1_{\ell,n}$}&
\multicolumn{1}{c}{$S^2_{\ell,n}$}&
\multicolumn{1}{c}{$T^1_{\ell,n}$}&
\multicolumn{1}{c}{$T^2_{\ell,n}$}\\
\hline
&\multicolumn{6}{l}{$\ell=1$}\\
 1 &   1.5861677 &  0.47187464 &  0.0027 & -0.1215 &  0.0742 & \multicolumn{1}{c}{--} \\
 2 &   1.1338905 &  0.46515116 &  0.1946 & -0.0991 &  0.1255 & \multicolumn{1}{c}{--} \\
 3 &   0.8807569 &  0.46565368 &  0.3473 & -0.0779 &  0.1769 & \multicolumn{1}{c}{--} \\
 4 &   0.7195665 &  0.46890412 &  0.4801 & -0.0604 &  0.2383 & \multicolumn{1}{c}{--} \\
 5 &   0.6082150 &  0.47259900 &  0.6021 & -0.0458 &  0.3131 & \multicolumn{1}{c}{--} \\
 6 &   0.5267854 &  0.47600569 &  0.7176 & -0.0330 &  0.4016 & \multicolumn{1}{c}{--} \\
 7 &   0.4646791 &  0.47895730 &  0.8289 & -0.0215 &  0.5037 & \multicolumn{1}{c}{--} \\
 8 &   0.4157567 &  0.48146486 &  0.9374 & -0.0108 &  0.6188 & \multicolumn{1}{c}{--} \\
 9 &   0.3762235 &  0.48358671 &  1.0440 & -0.0007 &  0.7469 & \multicolumn{1}{c}{--} \\
10 &   0.3436109 &  0.48538634 &  1.1490 &  0.0089 &  0.8877 & \multicolumn{1}{c}{--} \\
11 &   0.3162449 &  0.48692017 &  1.2530 &  0.0182 &  1.0411 & \multicolumn{1}{c}{--} \\
12 &   0.2929509 &  0.48823507 &  1.3562 &  0.0272 &  1.2069 & \multicolumn{1}{c}{--} \\
13 &   0.2728805 &  0.48936906 &  1.4586 &  0.0360 &  1.3852 & \multicolumn{1}{c}{--} \\
14 &   0.2554057 &  0.49035278 &  1.5606 &  0.0445 &  1.5758 & \multicolumn{1}{c}{--} \\
\hline
&\multicolumn{6}{l}{$\ell=2$}\\
 1 &   2.2168837 &  0.16413695 & -0.0603 & -0.1283 &  0.1162 &  0.0019 \\
 2 &   1.6817109 &  0.13416727 &  0.2040 & -0.1171 &  0.1492 & -0.0186 \\
 3 &   1.3499152 &  0.13379919 &  0.3992 & -0.1267 &  0.2263 & -0.0409 \\
 4 &   1.1271730 &  0.13720207 &  0.5686 & -0.1417 &  0.3295 & -0.0666 \\
 5 &   0.9676634 &  0.14084985 &  0.7244 & -0.1587 &  0.4555 & -0.0964 \\
 6 &   0.8478758 &  0.14409412 &  0.8717 & -0.1767 &  0.6031 & -0.1307 \\
 7 &   0.7546269 &  0.14685697 &  1.0136 & -0.1952 &  0.7716 & -0.1695 \\
 8 &   0.6799744 &  0.14918725 &  1.1517 & -0.2140 &  0.9608 & -0.2128 \\
 9 &   0.6188542 &  0.15115489 &  1.2870 & -0.2330 &  1.1704 & -0.2607 \\
10 &   0.5678867 &  0.15282450 &  1.4201 & -0.2521 &  1.4006 & -0.3132 \\
11 &   0.5247312 &  0.15424993 &  1.5517 & -0.2713 &  1.6500 & -0.3700 \\
12 &   0.4877153 &  0.15547464 &  1.6819 & -0.2905 &  1.9198 & -0.4315 \\
13 &   0.4556123 &  0.15653342 &  1.8111 & -0.3098 &  2.2095 & -0.4975 \\
14 &   0.4275023 &  0.15745412 &  1.9395 & -0.3291 &  2.5190 & -0.5679 \\
15 &   0.4026818 &  0.15825915 &  2.0672 & -0.3484 &  2.8482 & -0.6429 \\
16 &   0.3806035 &  0.15896664 &  2.1943 & -0.3678 &  3.1970 & -0.7223 \\
17 &   0.3608352 &  0.15959137 &  2.3209 & -0.3872 &  3.5655 & -0.8062 \\
18 &   0.3430314 &  0.16014547 &  2.4470 & -0.4066 &  3.9536 & -0.8945 \\
19 &   0.3269120 &  0.16063895 &  2.5728 & -0.4259 &  4.3613 & -0.9873 \\
20 &   0.3122481 &  0.16108017 &  2.6983 & -0.4453 &  4.7884 & -1.0846 \\
21 &   0.2988504 &  0.16147607 &  2.8234 & -0.4648 &  5.2351 & -1.1863 \\
22 &   0.2865611 &  0.16183254 &  2.9484 & -0.4842 &  5.7013 & -1.2924 \\
23 &   0.2752477 &  0.16215452 &  3.0731 & -0.5036 &  6.1869 & -1.4030 \\
24 &   0.2647981 &  0.16244623 &  3.1976 & -0.5230 &  6.6920 & -1.5180 \\
25 &   0.2551166 &  0.16271128 &  3.3220 & -0.5424 &  7.2164 & -1.6374 \\
\hline
&\multicolumn{6}{l}{$\ell=3$}\\
 1 &   2.6013404 &  0.06527813 & -0.1840 & -0.0702 &  0.0898 &  0.0117 \\
 2 &   2.0582624 &  0.04834015 &  0.0662 & -0.0551 &  0.0478 &  0.0014 \\
 3 &   1.6990205 &  0.05125011 &  0.2366 & -0.0576 &  0.0593 & -0.0023 \\
 4 &   1.4466219 &  0.05532661 &  0.3818 & -0.0631 &  0.0803 & -0.0054 \\
 5 &   1.2597371 &  0.05898678 &  0.5132 & -0.0695 &  0.1071 & -0.0085 \\
 6 &   1.1157943 &  0.06207639 &  0.6359 & -0.0764 &  0.1389 & -0.0120 \\
 7 &   1.0015072 &  0.06465872 &  0.7527 & -0.0836 &  0.1753 & -0.0158 \\
 8 &   0.9085566 &  0.06682318 &  0.8653 & -0.0909 &  0.2162 & -0.0200 \\
 9 &   0.8314693 &  0.06864897 &  0.9747 & -0.0983 &  0.2616 & -0.0246 \\
10 &   0.7664974 &  0.07020016 &  1.0817 & -0.1058 &  0.3113 & -0.0296 \\
11 &   0.7109883 &  0.07152736 &  1.1869 & -0.1133 &  0.3652 & -0.0350 \\
12 &   0.6630115 &  0.07267052 &  1.2906 & -0.1208 &  0.4235 & -0.0409 \\
13 &   0.6211284 &  0.07366127 &  1.3931 & -0.1284 &  0.4859 & -0.0471 \\
14 &   0.5842454 &  0.07452488 &  1.4946 & -0.1360 &  0.5526 & -0.0537 \\
15 &   0.5515158 &  0.07528169 &  1.5953 & -0.1437 &  0.6234 & -0.0608 \\
16 &   0.5222741 &  0.07594820 &  1.6953 & -0.1513 &  0.6985 & -0.0683 \\
17 &   0.4959899 &  0.07653788 &  1.7947 & -0.1590 &  0.7776 & -0.0761 \\
18 &   0.4722351 &  0.07706183 &  1.8936 & -0.1666 &  0.8609 & -0.0844 \\
19 &   0.4506608 &  0.07752926 &  1.9920 & -0.1743 &  0.9483 & -0.0931 \\
20 &   0.4309792 &  0.07794783 &  2.0901 & -0.1820 &  1.0399 & -0.1022 \\
21 &   0.4129512 &  0.07832398 &  2.1878 & -0.1897 &  1.1356 & -0.1117 \\
22 &   0.3963764 &  0.07866313 &  2.2853 & -0.1974 &  1.2353 & -0.1216 \\
23 &   0.3810855 &  0.07896988 &  2.3825 & -0.2051 &  1.3392 & -0.1320 \\
24 &   0.3669346 &  0.07924815 &  2.4794 & -0.2128 &  1.4472 & -0.1427 \\
25 &   0.3538006 &  0.07950128 &  2.5762 & -0.2205 &  1.5592 & -0.1538 \\
\hline

  \end{tabular}
\end{table*}

To know the coefficients for another normalization, for instance for $\tilde\omega=\omega/\Omega_K$, one can use the following development:
\begin{equation}
\tilde\Omega= \frac{\Omega}{\Omega_K}= \bar\Omega + A \bar\Omega^3 + {\cal O}(\bar\Omega^5).
\end{equation}
The perturbed frequencies in this new normalization then express
\begin{multline}
\tilde\omega_{\ell,m,n}^{pert}=\bar\omega_{\ell,n}^{(0)}+ mC_{\ell,n} \tilde\Omega + 
(S^1_{\ell,n}+A\bar\omega_{\ell,n}^{(0)}+m^2S^2_{\ell,n}) \tilde\Omega^2 + \\
m(T^1_{\ell,n}+m^2T^2_{\ell,n}) \tilde\Omega^3 + {\cal O}(\tilde\Omega^4). \label{eq:pertfin2}
\end{multline}
From our models we have computed $A \approx 0.77164$.

\subsection{Coefficient accuracy and comparisons with previous works}\label{ssec:accurcoef}

The zeroth-order coefficients $\bar\omega^{(0)}_{\ell,n}$ were compared to the 1-D computations and we find agreement within $10^{-9}$. We also compared our results to previous frequency computations of in  a nonrotating polytropic model performed by \citet{CDM94} with a totally different method. We renormalized their results for g modes (Table 4 of their paper) to their dynamical frequency $\nu_g=99.8557\unit{\mu Hz}$ (Eq. 3.2 of their paper). The relative differences with our results do not exceed $5\times 10^{-8}$.

The choice for the step $\delta\bar\Omega$ is important for the accuracy of the terms $C_{\ell,m,n}^j$. Ideally, we should choose $\delta\bar\Omega$ as small as possible, but when it is too small, the numerical noise, produced by the uncertainties on the computed $\omega_{\ell,m,n}$ (Sect.~\ref{ssec:resol}), drastically increases. We then chose the value of $\delta\bar\Omega$ to have the best trade-off. These uncertainties on $C_{\ell,m,n}^j$ determinations were taken into account for the estimated accuracy of the coefficients $C_{\ell,n}$, $S^i_{\ell,n}$ and $T^i_{\ell,n}$.

The 1st-order perturbative coefficients $C_{\ell,n}$ are expressed with integrals of the eigenmodes in the nonrotating model \citep{Ledoux51}. We then computed these terms with our 1-D eigensolutions and compared them to $C_{\ell,m,n}^1/m$. The results are consistent within $10^{-8}$.

An explicit computation of 2nd- and 3rd-order coefficients requires calculating the 1st- and 2nd-order corrections of the eigenfunctions, which is not so straightforward. It is the reason we performed a direct numerical computation of these coefficients. The numerical errors we estimated for $S^i_{\ell,n}$ and $T^i_{\ell,n}$ are generally around $10^{-5}$ and always less than $10^{-4}$. We then checked the consistency of our computations with the 2nd-order calculations of \citet{Saio81} for g modes with $n=1$ to 3. In this work, all frequencies were normalized by the dynamical frequency $\Omega_K^{(0)}$ of the nonrotating polytrope. By noticing that $\Omega/\Omega_K^{(0)}=\bar\Omega + A\bar\Omega^3+{\cal O}(\bar\Omega^5)$ with $A\approx0.18391$, and using the relation~(\ref{eq:pertfin2}), 
we were able to compare these results with ours. We get a good qualitative agreement with absolute differences better than $10^{-2}$, which is reasonable relative to the lower accuracy of Saio's computations. It gives an interesting consistency check for our calculations.
Overall, the perturbative coefficients listed in Table~\ref{tab:coef} have been determined with high accuracy.

\section{Domains of validity of perturbative approaches}\label{Sec:Validity}

\begin{figure}[htp]
  \centering
  \includegraphics[width=\linewidth]{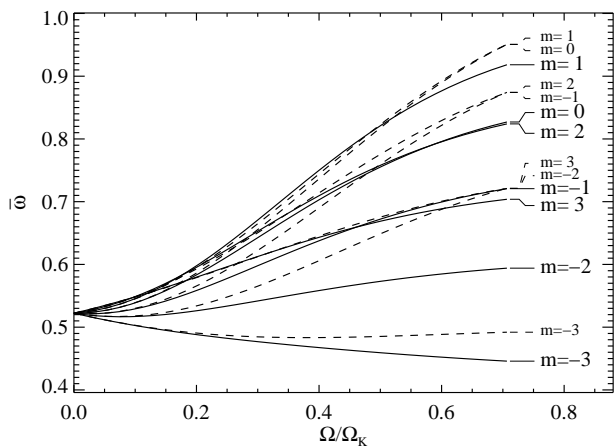}
  \caption{Evolution with the rotation rate of the frequencies of the components of the ($\ell=3$, $n=16$) multiplet obtained with a complete computation (solid line) and with the  2nd-order perturbative approximation (dashed lines).}
  \label{fig:l3n16}
\end{figure}

From the previously computed coefficients, we calculated mode frequencies with the 1st to 3rd-order perturbative approximations for rotation rates ranging from $\Omega=0$ to $0.7\Omega_K$ and compared them to complete computations.
Figure~\ref{fig:l3n16} illustrates such a comparison by showing the evolution of the frequencies of the seven $m$ components of an $\ell=3$ mode together with their 2nd-order perturbative approximation. We clearly observe that the agreement between both approaches at low rotation progressively disappears as the rotation increases.

To define the domains of validity of perturbative approaches, we fix the maximal departure $\delta\bar\omega$ allowed between the perturbed frequencies $\bar\omega^{pert}_{\ell,m,n}$ and the ``exact'' ones $\bar\omega_{\ell,m,n}$. 
For each mode and each approximation order, we define the domain of validity $[0,\Omega_v]$, such that $|\bar\omega^{pert}_{\ell,m,n}(\Omega)- \bar\omega_{\ell,m,n}(\Omega)| < \delta\bar\omega\ \forall \Omega<\Omega_v$. 
The precision of the observed frequencies $\delta \nu$ can be related to the normalized error $\delta\bar\omega$ through
\begin{equation}
 \label{eq:errordim}
 \delta\bar\omega=2\pi\,\delta\nu\sqrt{\frac{R^3}{GM}}.
\end{equation}
From this expression, we see that, for a fixed precision $\delta \nu$, the normalized error $\delta\bar\omega$
depends on the dynamical frequency $\nu_g= (GM/R^3)^{\frac{1}{2}}/(2\pi)$ of the star considered.
We thus display the domains of validity of the perturbative approximations for two types of stars
with different dynamical frequencies,
a typical $\gamma$~Dor star, and a typical B star.
The $\gamma$~Dor star is such that $M=1.55M_{\sun}$, $R=1.6R_{\sun}$, i.e. $\nu_g=61\unit{\mu Hz}$, while the B star has $M=4M_{\sun}$, $R=7R_{\sun}$, i.e. $\nu_g=11\unit{\mu Hz}$.
For the frequency precision, we chose $\delta\nu=0.1\unit{\mu Hz}$, which corresponds to the resolution of an oscillation spectrum after a hundred days. This is the typical accuracy for a CoRoT long run.
Accordingly, the normalized error $\delta\bar\omega$ is equal to $1.6\times 10^{-3}$ for the $\gamma$~Dor star
and $9.3\times 10^{-3}$ for the B star.
It must be noted that the domains of validity 
will not be affected by the numerical errors of $\bar\omega^{pert}_{\ell,m,n}$ (cf.~Sect.~\ref{ssec:accurcoef}) and $\bar\omega_{\ell,m,n}$ (cf.~Sect.~\ref{ssec:resol})
as the values of $\delta\bar\omega$ remain at least an order of magnitude higher. 

\begin{figure*}[!htp]
  \centering
  \includegraphics[width=.5\linewidth]{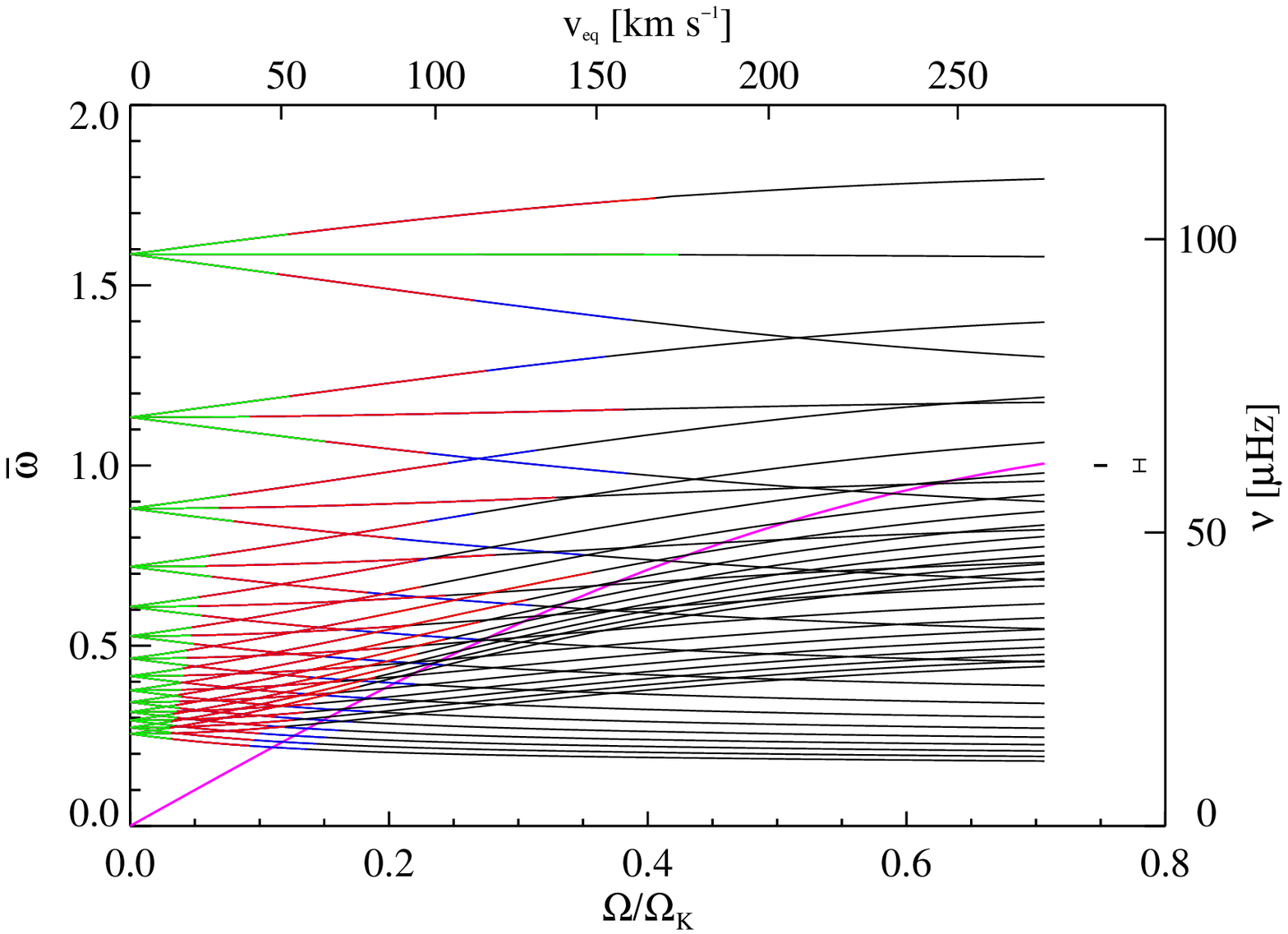}\includegraphics[width=.5\linewidth]{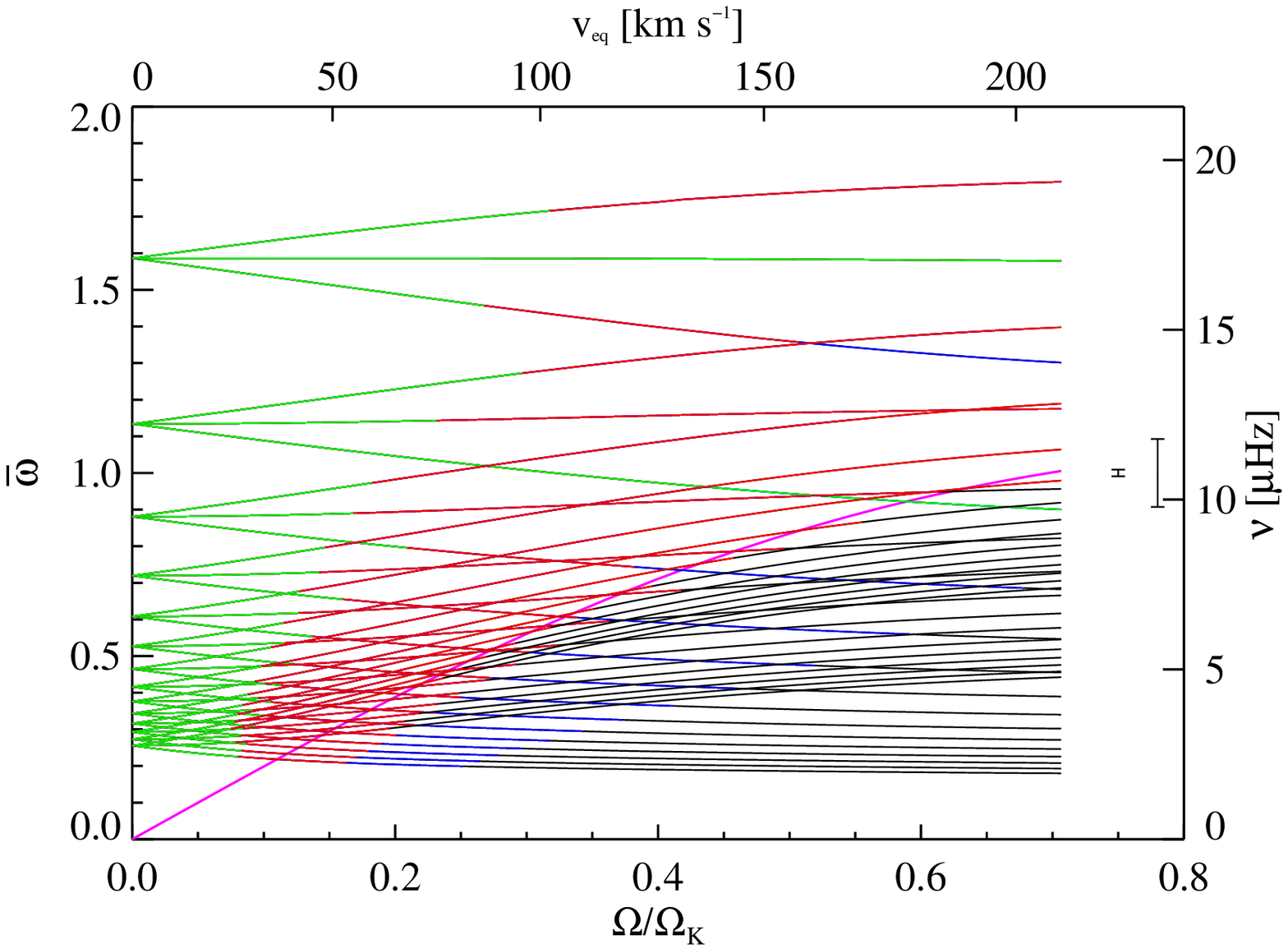}\\
  \includegraphics[width=.5\linewidth]{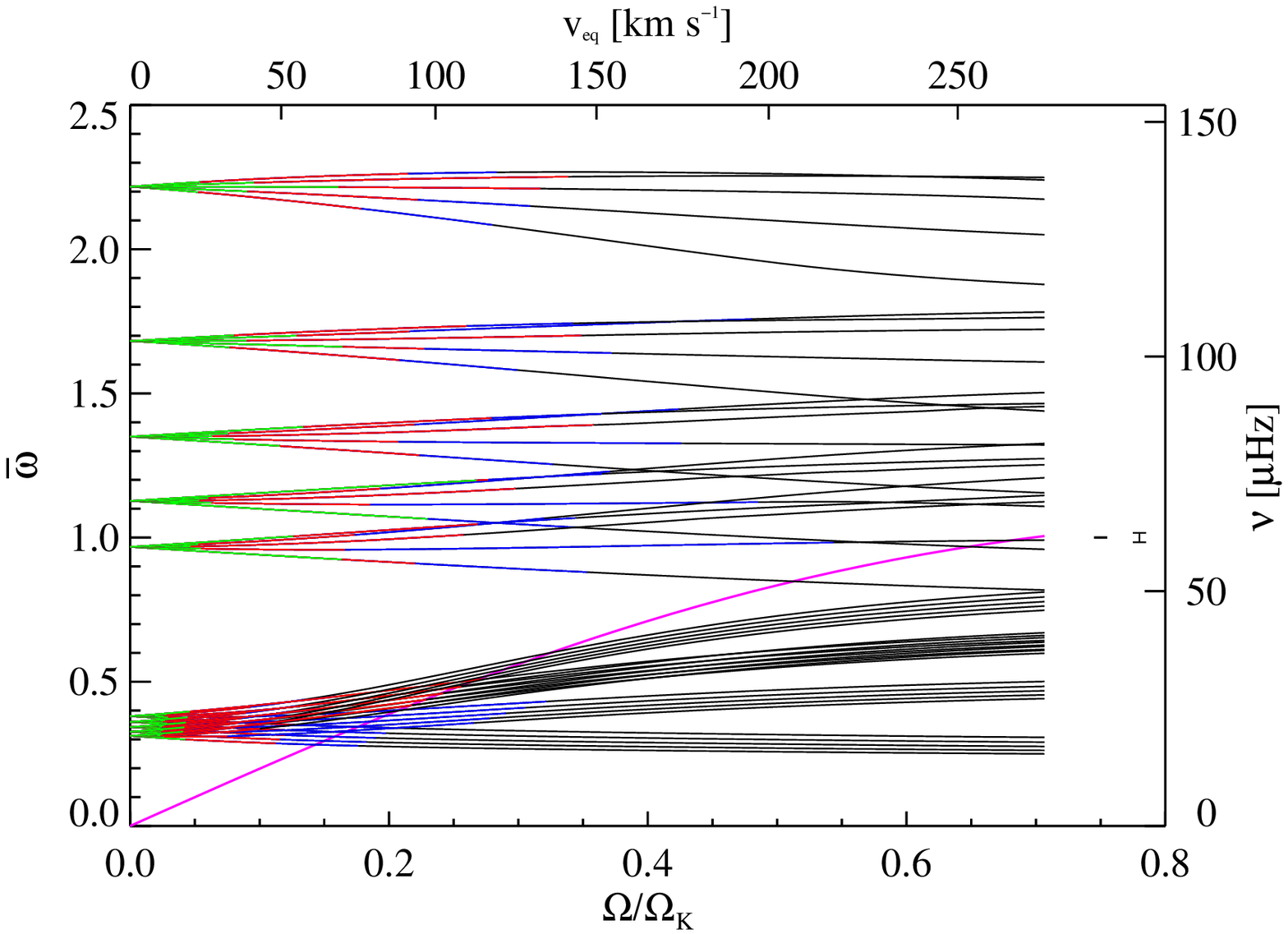}\includegraphics[width=.5\linewidth]{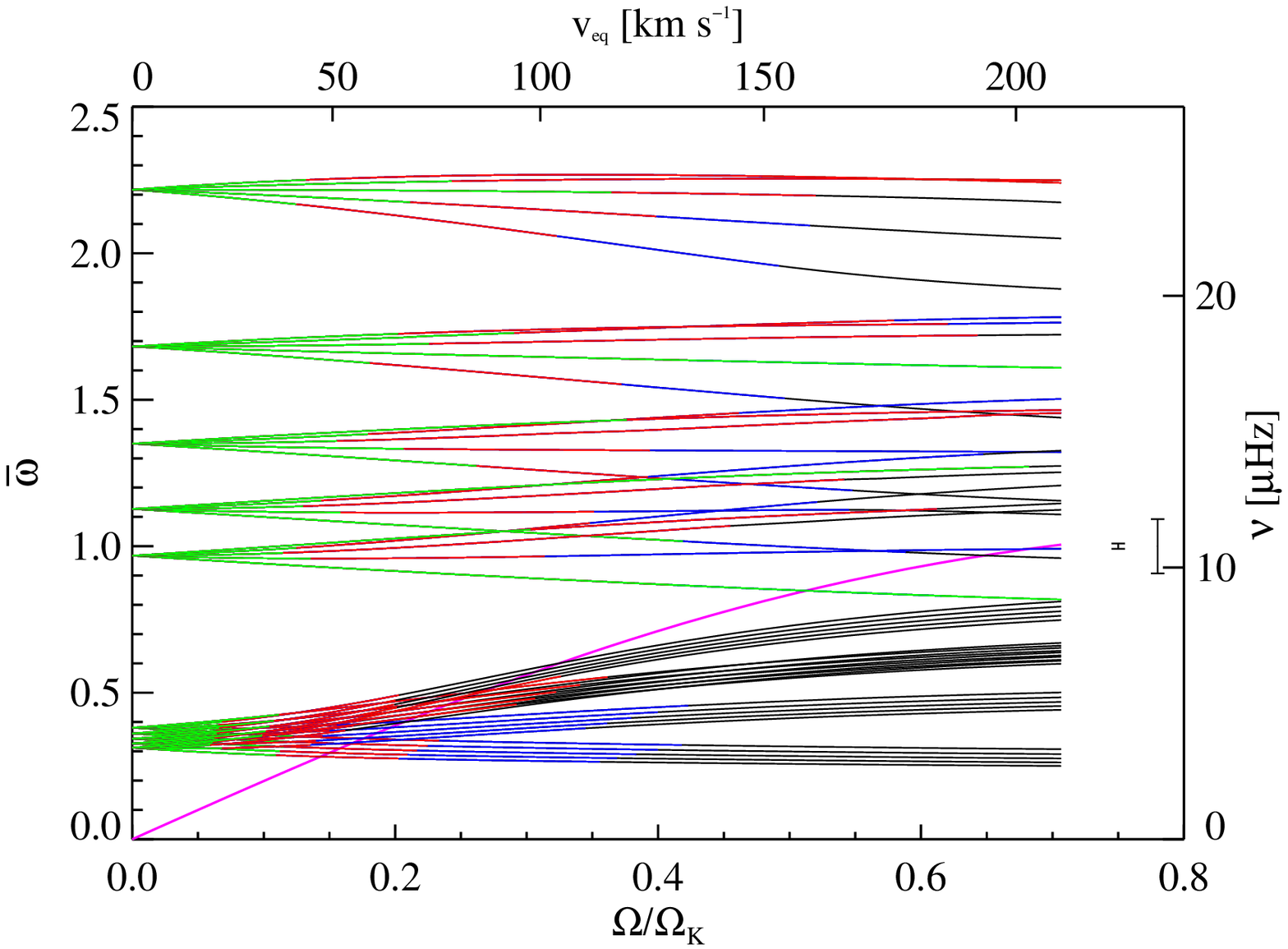}\\
  \includegraphics[width=.5\linewidth]{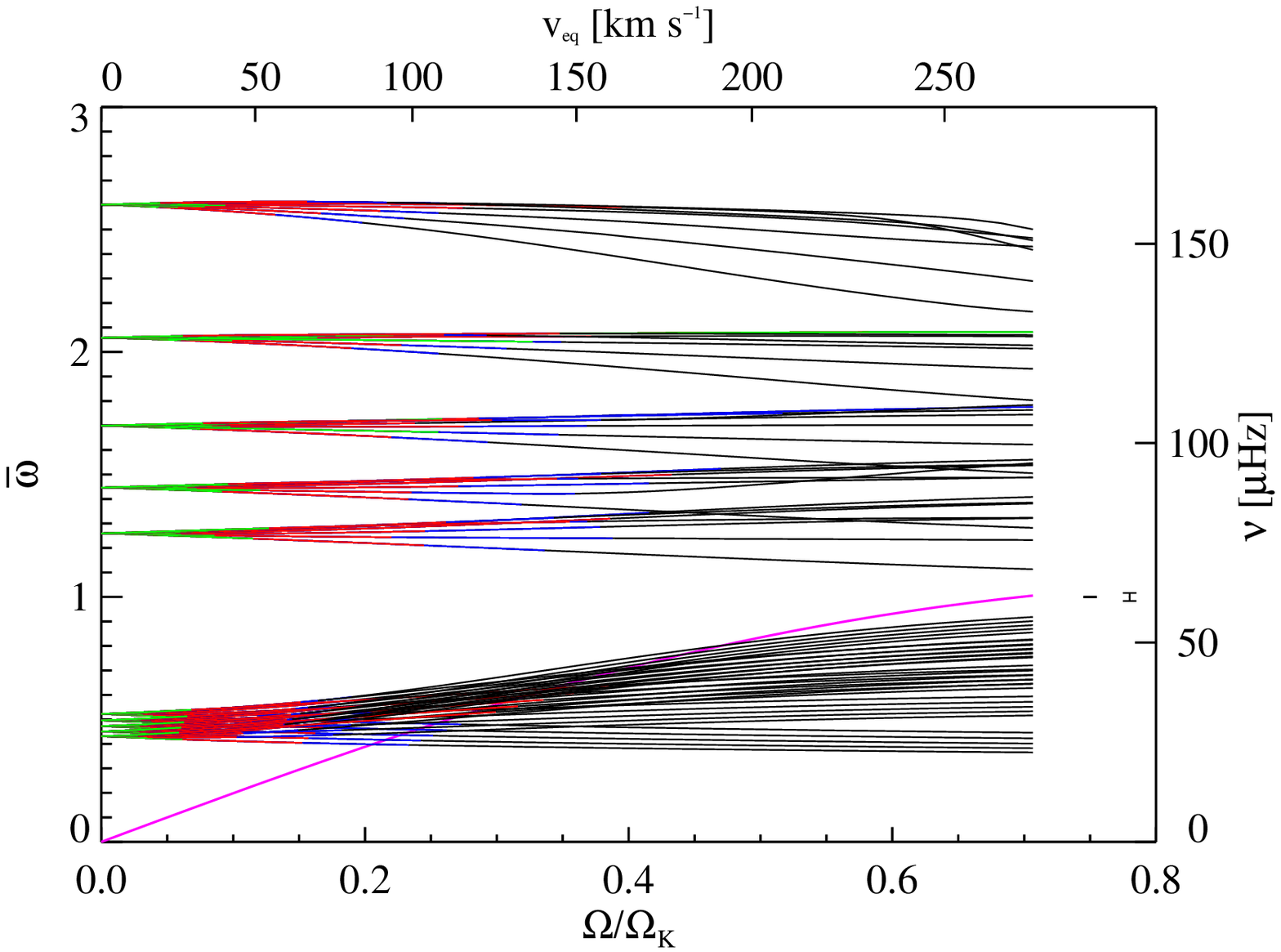}\includegraphics[width=.5\linewidth]{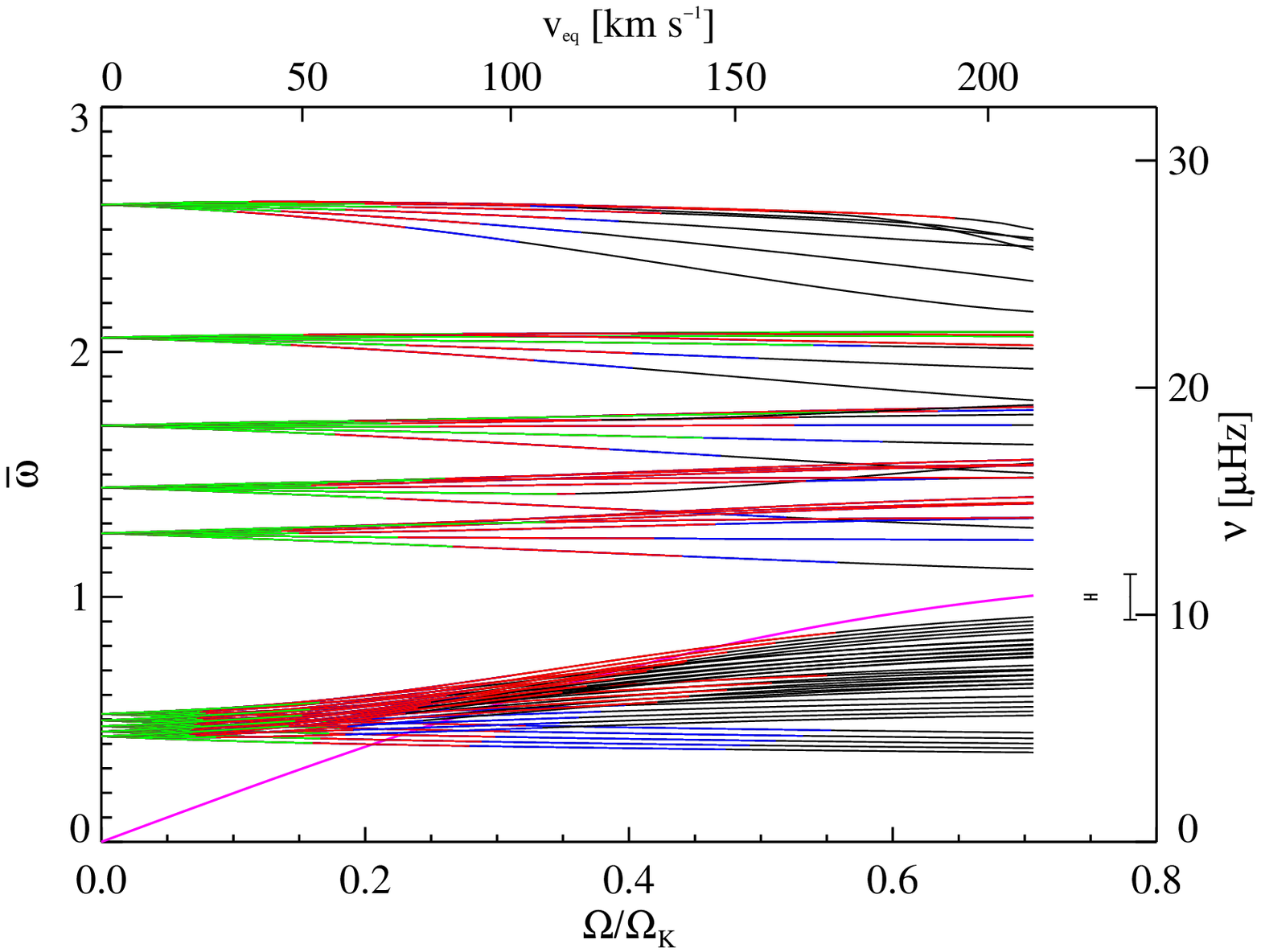}
  \caption{Evolution of the frequencies of $\ell=1,2,3$ modes \textit{(top to bottom)}. Frequencies are computed in the corotating frame. 
Perturbative approximations have been tested for a typical $\gamma$~Dor \textit{(left panels)} 
and for a B star \textit{(right panels)}. Green/red/blue parts of curves indicate that 1st/2nd/3rd order is sufficient to reproduce complete calculations within an error $\delta\nu =0.1\unit{\mu Hz}$. Error bars on the righthand side of each panel show $\delta\nu$ and $10\times\delta\nu$. Magenta lines indicate $\omega=2\Omega$. For each plot, the bottom x-axis and left y-axis show dimensionless units, whereas the top x-axis and right y-axis show physical units.}
  \label{fig:valid}
\end{figure*}

We have determined the domains of validity of 1st-, 2nd-, and 3rd-order methods for low-degree $\ell \le 3$ modes. Specifically,
we considered $\ell=1$ modes with $n=1$ to 14, and $\ell=2$ and 3 low-order ($n=1$ to 5), and high-order ($n=16$ to 20) modes.
The domains of validity are shown in Fig.~\ref{fig:valid} for both types of stars.
Overall, the domains of validity extend to higher rotation rates for B stars than for $\gamma$~Dor stars.
This is simply due to the increase in the normalized
tolerance $\delta\bar\omega$.
Besides, we observe distinct behaviors in the high- and low-frequency ranges.

In the high-frequency range, 2nd-order perturbative methods give satisfactory results up to $\sim$100\unit{km\,s^{-1}} for $\gamma$~Dor stars
and up to $\sim$150\unit{km\,s^{-1}} for B stars. The 3rd-order terms improve the results and increase the domains of validity by a few tens of
\unit{km\,s^{-1}}. These results are to be contrasted with those found for p modes where
the domains of validity are restricted to lower rotation rates. For $\delta$ Scuti stars, which have similar stellar parameters to $\gamma$~Dor,
\citet{Reese06} find $\sim$50--70\unit{km\,s^{-1}} as a limit for perturbative methods. In addition, the 3rd-order terms do
not improve the perturbative approximation in this case, as p modes are weakly sensitive to the Coriolis force.
The rather good performance of perturbative methods at describing high-frequency g modes 
indicates in particular that the 2nd-order term gives a reasonable description of the centrifugal distortion.
This might be surprising considering the significant distortion of the stellar surface ($R_{eq}/R_p = 1.08$ at $\Omega = 0.4 \Omega_K$). 
Actually, the energy of g modes is concentrated in the inner part of the star where 
the deviations from sphericity remain small (as shown in Fig.~\ref{fig:BV}-\textit{top}).  As a result, g modes ``detect'' a much weaker distortion that is
then amenable to a perturbative description.
A particular feature that induces a strong deviation from the perturbative method concerns
mixed pressure-gravity modes that arise as a consequence of the centrifugal modification of the stellar structure.
For example, we found that, above a certain rotation rate, the $\ell=3,n=1$ mode becomes a mixed mode with a p-mode character
in the outer low-latitude region associated with a drop in the Brunt-V\"ais\"al\"a frequency $N_o$ (see Fig.~\ref{fig:BV}). 

The domains of validity of perturbative methods are strongly reduced in the low-frequency range.
For $\gamma$Dor stars, 2nd-order perturbative methods are only valid below $\sim$50\unit{km\,s^{-1}}.
Indeed, a striking feature of Fig.~\ref{fig:valid} is that perturbative methods fail to recover 
the correct frequencies in the inertial regime $\omega<2\Omega$ (delimited by a magenta curve).
In particular, we observe that, although increasing the tolerance $\delta\bar\omega$ between
the left ($\gamma$~Dor) and right (B star) panels subtantially extends the domains of validity in the $\omega > 2 \Omega$ regime,
very little improvement is observed in the $\omega<2\Omega$ regime.

\begin{figure}[!tp]
  \centering
  \includegraphics[height=.44\linewidth]{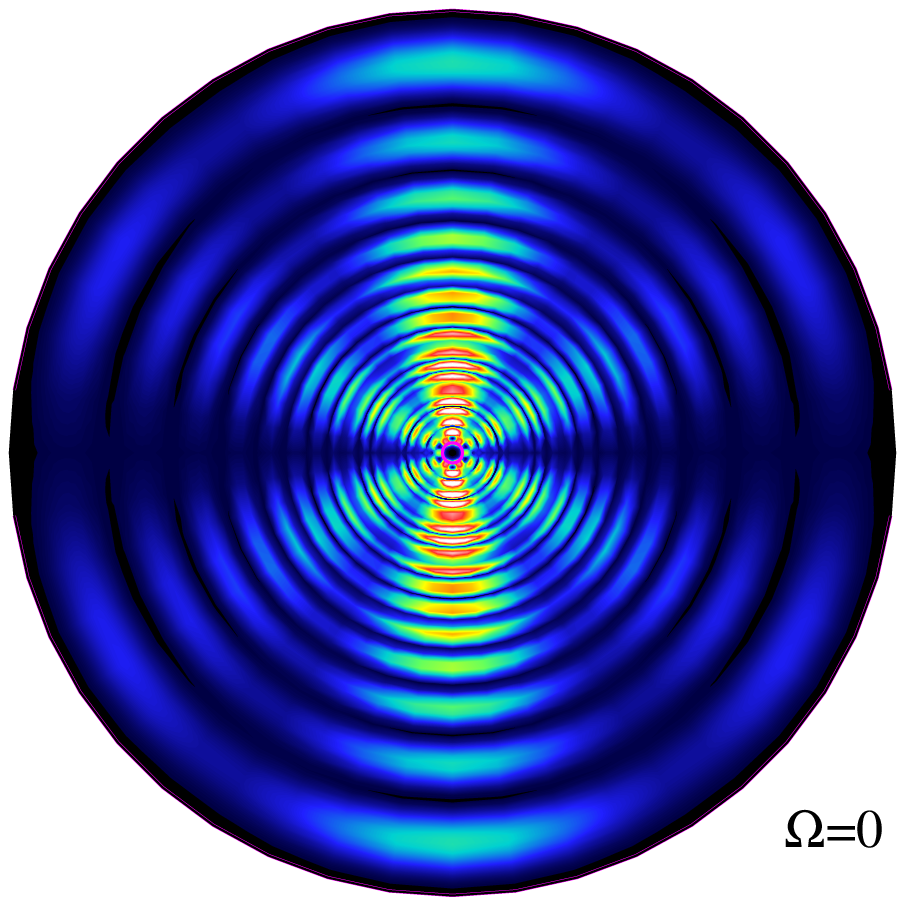}%
  \includegraphics[height=.44\linewidth]{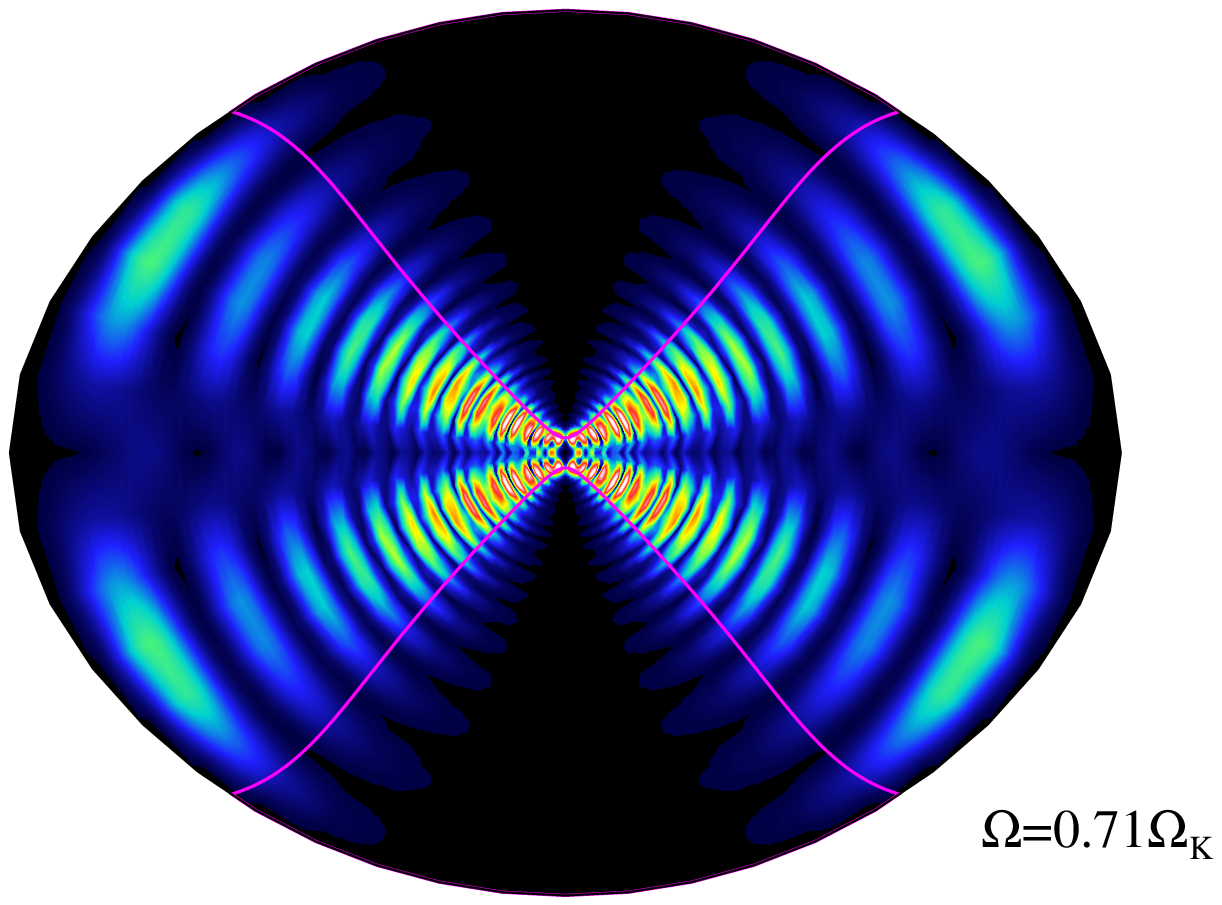}
  \caption{\textit{(Left)} Meridional distribution of kinetic energy  $\frac{1}{2}\rho_o\vec v^2$ of the g mode $(\ell=3,m=-1,n=16)$ in a
nonrotating star. To enhance the contrast, it is scaled by $r^2$. \textit{(Right)} The same for $\Omega=0.7\Omega_K$. Magenta lines indicate the critical surface $\Gamma=0$.}
  \label{fig:forbid}
\end{figure}

In the following, we argue that the failure of the perturbative method in the subinertial regime $[0,2 \Omega]$
is related to changes in the mode cavity that are not taken into account 
by the perturbative method.
Indeed,
we observed that modes in the inertial regime do not explore the polar region 
and that the angular size of this forbidden region increases with $2\Omega/\omega$. This is illustrated in Fig.~\ref{fig:forbid}
for a particular mode. Such a drastic change
in the shape of the resonant cavity has a direct impact on the associated mode frequency.
As perturbative methods totally ignore this effect, they cannot provide an accurate approximation of the frequencies in this regime.

This interpretation is supported by the analytical expression of the forbidden region determined by \citet{Dintrans00} for gravito-inertial modes. Indeed for frequencies $\omega < 2\Omega$, modes are mixed gravity-inertial modes, since the Coriolis force becomes a restoring force. In the context of their spherical model, and within the anelastic approximation and the Cowling approximation, they have shown that gravito-inertial waves with a frequency $\omega$ only  propagate in the region where
\begin{equation}
\Gamma = r^2 \omega^2 [N_o^2 +(2\Omega)^2 -\omega^2]
- (2\Omega N_o z)^2 > 0. \label{eq:crit}
\end{equation}
This implies that, when $\omega<2\Omega$, a critical latitude $\theta_c=\arcsin[\omega/(2\Omega)]$ appears above which waves cannot propagate.
Even though this expression does not strictly apply to our nonspherical geometry, we have overplotted 
the critical surfaces $\Gamma=0$ with the energy distributions of our eigenmodes (see Fig.~\ref{fig:forbid} for an illustration).
For the nonrotating case, there is only a small circle close to the center, corresponding to the classical turning point $\omega=N_o$. For the mode with $\omega<2\Omega$, the polar forbidden region delineated by $\Gamma=0$ agrees pretty well with the energy distribution of our complete computation.

\section{Conclusion}\label{Sec:Concl}

In the present work, we have computed accurate frequencies for g modes in polytropic models of uniformly spinning stars.
We started from high- and low-frequency, low-degree ($\ell \le 3$) g modes of a nonrotating star and followed them up to $\Omega = 0.7 \Omega_K$.
This allowed us to provide a table of numerically-computed perturbative coefficients up to the 3rd order for a polytropic stellar structure (with index $\mu=3$). This table can serve as a reference for testing the implementation of perturbative methods.
We then determined the domains of validity of perturbative approximations. For the high-frequency (low-order) modes, 2nd-order perturbative methods correctly describe modes up to $\sim 100\unit{km\,s^{-1}}$ for $\gamma$~Dor stars and up to $\sim 150\unit{km\,s^{-1}}$ for B stars. The domains of validity can be extended by a few tens of \unit{km\,s^{-1}} with 3rd-order terms. However, the domains of validity shrink at low frequency. In particular, perturbative methods fail in the inertial domain $\omega < 2\Omega$ because of a modification in the shape of the resonant cavity.

In a next step, we plan to compare our complete computations with the so-called traditional approximation, which is also extensively used to determine g-mode
frequencies \citep[e.g][]{Berthomieu78,LeeSaio97}. We will also analyze how rotation affects the regularities of the spectrum  -- such as the period spacing -- and compare it to the predictions of the perturbative and traditional methods. In the present study, we have focused on low-degree modes, but a more complete exploration clearly needs to be performed. In particular, we might look for the singular modes predicted by \citet{Dintrans00}. It requires to take care of dissipative processes, which play an important role in this case.

\begin{acknowledgements}
The authors acknowledges support through the ANR project Siroco. Many of the numerical calculations were carried out on the supercomputing facilities of CALMIP (``CALcul en MIdi-Pyr\'en\'ees''), which is gratefully acknowledged. The authors also warmly thank Boris Dintrans for discussions and useful comments on this work. DRR gratefully acknowledges support from the CNES (``Centre National d'{\'E}tudes Spatiales'') through a postdoctoral fellowship.
\end{acknowledgements}

\bibliographystyle{aa}
\bibliography{biblio}

\end{document}